\documentclass[preprint,1p,leqno]{elsarticle}
\usepackage{amsmath}
\usepackage{amssymb}
\usepackage{mathrsfs}
\usepackage{color}
\usepackage{marvosym}

\usepackage{lineno}
\modulolinenumbers[1]

\usepackage[]{graphicx}
\usepackage{float}
\usepackage{psfrag}
\usepackage{pdfpages}
\usepackage[dvipsnames]{xcolor}
\usepackage[makeroom]{cancel}

\usepackage{amsthm}

\usepackage{bigstrut}
\usepackage{multirow}
\usepackage{subfig}

\usepackage{hyperref}

\usepackage{geometry}
\journal{arXiv}

\usepackage[linesnumbered,ruled,vlined]{algorithm2e}

\begin{document}
\begin{frontmatter}

\title{A moving least square immersed boundary method for SPH with thin-walled structures}

\author[SYSU]{ZhuoLin Wang}
\author[SYSU]{Zichao Jiang}
\author[TU]{Yi Zhang}
\author[SYSU]{Gengchao Yang}
\author[SYSU]{Trevor Hocksun Kwan}
\author[SYSU]{Yuhui Chen}
\author[SYSU]{Qinghe Yao \corref{cor1}}

\ead{yaoqh@mail.sysu.edu.cn}
\address[SYSU]{Sun Yat-sen University, School of Aeronautics and Astronautics, Shenzhen, China}
\address[TU]{ University of Twente, Robotics and Mechatronics, Enschede, The Netherlands}
\cortext[cor1]{Corresponding author}

\begin{abstract}
This paper presents a novel method for smoothed particle hydrodynamics (SPH) with thin-walled structures. Inspired by the direct forcing immersed boundary method, this method employs a moving least square method to guarantee the smoothness of velocity near the structure surface. It simplifies thin-walled structure simulations by eliminating the need for multiple layers of boundary particles, and improves computational accuracy and stability in three-dimensional scenarios. Supportive three-dimensional numerical results are provided, including the impulsively started plate and the flow past a cylinder. Results of the impulsively started test demonstrate that the proposed method obtains smooth velocity and pressure in the, as well as a good match to the references results of the vortex wake development. In addition, results of the flow past cylinder test show that the proposed method avoids mutual interference on both side of the boundary, remains stable for three-dimensional simulations while accurately calculating the forces acting on structure.

\end{abstract}
\end{frontmatter}


\section{Introduction}

Over the past decades, meshless methods have undergone remarkable advancements \cite{luo2021particle}. The availability of various high-performance algorithms \cite{springel2005cosmological, yao2023towards} has facilitated their widespread applications by alleviating concerns regarding computational complexity. Among these methods, smoothed particle hydrodynamics (SPH) \cite{monaghan1992smoothed}, due to its natural adaptability to large-deformations and moving boundaries, shows a good potential in free-surface flows, multi-phase flows \cite{wang2016overview,nair2019simulations}, fluid-structure interaction (FSI) problems \cite{sun2021accurate,xianpeng2023improved,zhang2021multi}, etc. Techniques like particle shifting \cite{khayyer2016comparative,lind2012incompressible,huang2019kernel} and various diffusion methods \cite{antuono2012numerical,antuono2010free} have been employed to reduce the instability of SPH caused by irregular particle distribution. In 2017, Sun et al. proposed the $\delta^+$-SPH \cite{sun2017deltaplus} method. It combines the shifting method with $\delta$-SPH \cite{marrone2011delta,bouscasse2013nonlinear} and achieves excellent results in reducing the tensile instability and the pressure oscillation, which is a solid foundation of the current study.

The treatment of fluid-structure boundaries stands as a pivotal aspect in SPH. Due to the reliance of various computations on the kernel approximation \cite{liu2010smoothed}, the treatment of fluid-structure boundaries differs from that in grid-based methods. Since the kernel approximation in the SPH method is derived under the assumption of a complete integral domain, the treatment of fluid-structure boundaries typically requires maintaining the integrity of the domain. A classic approach is putting some particles at the structure surface to fill the integral domain. For example, in 1997, Morris et al. simulated the solid wall by placing a series of fixed particles along the solid wall and computed the velocity of those particles according to the tangent plane and the distance to the boundary surface \cite{morris1997modeling}. In 2003, by directly replicating a series of mirrored particles along the structure surface, Colagrossi et al. successfully simulated both slip and no-slip boundaries \cite{colagrossi2003numerical}. Commonly used extensions of this approach are proposed by Adami et al. \cite{adami2012generalized} and Marrone et al. \cite{marrone2011delta}; Adami et al. use the renormalized kernel approximation to compute velocity and pressure of the boundary particles, and Marrone et al. calculate velocity and pressure of the boundary particles by doing a kernel approximation at the mirrored position at the boundary surface.

Another approach is using the non-vanishing surface integral for the incomplete integral domain while deriving the kernel approximation. This approach does not need extra particles to fill the integral domain. Thus it is easier to model complex geometries. In 2012, Marrone et al. successfully applied this approach to ship wave breaking patterns \cite{marrone2012study}. In 2019, Chiron et al. further developed this approach by introducing a Laplacian operator and a cutface process for calculating the particle/wall interactions on any type of geometry \cite{chiron2019fast}.

For simulating problems with structures immersed in fluid, another commonly used technique is the immersed boundary method (IBM) \cite{peskin2002immersed}. Compared with the approaches introduced in previous paragraphs, the IBM requests neither multiple layers of boundary particles nor the calculation of the surface integral. Thus it can more easily incorporate thin-walled structures immersed in fluid.

In the IBM, a structure is represented by a series of Lagrange particles. These particles interact with the flow field by imposing boundary force to grid nodes nearby. There are different ways proposed to calculate this boundary force. The approach proposed by Peskin \cite{peskin1972flow} is to calculate it according to the deformation of the flexible structure. A second approach is the feed-back forcing scheme \cite{GOLDSTEIN1993354}, which introduces a feed-back mechanism based on the desired velocity and the actual velocity near the structure to compute the boundary force. Another approach is the direct forcing scheme \cite{fadlun2000combined, kajishima2002interaction}. It calculate the force by finding the difference between preliminary velocity and desired velocity, enforcing the velocity of fluid at the structure surface to be equal to the structure velocity at the end of a time step. An extension is the diffusive direct forcing IBM proposed by Uhlmann \cite{UHLMANN2005448} with the aim of reducing the oscillations caused by the velocity interpolation. In this extension, the boundary force is calculated on the structure particles and then is spread to fluid grid nodes using a Dirac delta function.

Some works have applied the IBM method in SPH. Hiber et al. \cite{hieber2008immersed} combined the remeshing SPH with the IBM for simulations of flows past complex deforming geometries in 2008. In the work of Kalateh et al., the IBM was used to couple SPH with finite element method \cite{kalateh2018application}. In 2019, it was also used to couple SPH with the discrete element method by Nasar et al. \cite{nasar2019flexible}.
These works introduce the widely used boundary force calculation approach in SPH; the boundary force is first calculated on the structure particles, and when spreading the force to fluid particles, the kernel function is then used as the Dirac delta function. However, the structure particles which consist of only one layer of points cannot fulfill the assumption that the integral domain is complete. This can lead to non-conservation of momentum \cite{NASAR2019263}. To avoid this, Nasar et al. \cite{NASAR2019263} and Cherfils \cite{cherfils2011developpements} introduced the moving least square (MLS) method to update the integral kernel when applying the diffusive direct forcing IBM in SPH. As will be explained in this paper, the use of the MLS method is equivalent to doing an extrapolation for the force on the structure surface. These works are for 2D problems. And when it comes to 3D problems, the extrapolation on a single layer of points may not be accurate and the coefficient matrix which must be inverted may become singular. This may lead to unstable problems. For more examples of using the IBM in SPH, we refer to, for instance, \cite{tan2020smoothed, ye2017hybrid, moballa2019dfib}.  

In this work, instead of the diffusive direct forcing approach, a direct forcing scheme is coupled with the MLS method for the velocity interpolation in  $\delta^+$-SPH. This choice enables us to achieve smoother results and to alleviate oscillations. In addition, it keeps stable in 3D simulations while calculating the forces acting on structure more accurately.
 
The organization of this paper is as follows: In Section~\ref{Discrete scheme}, a brief introduction to $\delta^+$-SPH is present. In Section~\ref{IBM}, the MLS diffusive direct forcing method and the proposed method are explained. Supportive numerical results follow in Section~\ref{result}. Finally conclusions are drawn in Section~\ref{Sec: conclusion}.

\section{$\delta^+$-SPH}\label{Discrete scheme}

The Naiver-Stokes equations for weakly compressible media can be written as
\begin{equation}\label{NS}
	\left\{
	\begin{aligned}
	&\frac{{\rm d}\boldsymbol{u}}{{\rm d}t}=-\frac{1}{\rho}\nabla p + \frac{\mu}{\rho}\nabla^2 \boldsymbol{u} + \frac{\boldsymbol{f}}{\rho}\\
	&\frac{{\rm d}\rho}{{\rm d}t}= -\rho \nabla\cdot\boldsymbol{u}
	\end{aligned}
	\right. , 
\end{equation}
which, together with the equation of the state, govern the dynamics of velocity $\boldsymbol{u}$, density $\rho$ and static pressure $p$ subject to external body force $\boldsymbol{f}$, material parameter dynamic fluid viscosity $\mu$, and initial and boundary conditions. 

The discrete form of \eqref{NS} in $\delta^+$-SPH \cite{sun2017deltaplus} is: 
\begin{equation}\label{NSdiscrete}
	\left\{
	\begin{aligned}
		&\frac{{\rm d}\boldsymbol{u}_i}{{\rm d}t}=\frac{1}{\rho_i}\sum_{j}F_{ij}\nabla_i W_{ij} V_j + K\frac{\mu}{\rho_i}\sum_{j}\pi_{ij}\nabla_i W_{ij} V_j + \frac{\boldsymbol{f}_i}{\rho_i}\\
		&\frac{{\rm d}\rho_i}{{\rm d}t}=-\rho_i\sum_{j}\left(\boldsymbol{u}_j-\boldsymbol{u}_i\right)\cdot\nabla_i W_{ij} V_j + \delta h c_0 \sum_{j}D_{ij}\cdot\nabla_i W_{ij}V_j \\
	\end{aligned}
	\right. ,
\end{equation}
where subscript $i$ represents the index of the particle, and the indices of
its neighboring particles within the supporting domain are denoted by subscript $j$. $V = m/\rho$ is the volume of particle. $h=1.3\Delta x$ is the smooth length. $\Delta x$ is the initial particle spacing.
$W_{ij}$ is the quintic spline kernel function,
\begin{equation*}
	W_{ij} = W(\left\lvert \boldsymbol{r}_{ij}\right\rvert, h) = W_0\times \left\{ 
	\begin{aligned}
		&\left(3-q\right)^5-6\left(2-q\right)^5+15\left(1-q\right)^5, &0\leqslant&q<1\\
		&\left(3-q\right)^5-6\left(2-q\right)^5, &1\leqslant&q<2\\
		&\left(3-q\right)^5, &2\leqslant&q<3\\
		&\ 0, &3\leqslant&q
	\end{aligned}
	\right. , 
\end{equation*}
where $\boldsymbol{r}_{ij} = \boldsymbol{r}_i - \boldsymbol{r}_j$ with $\boldsymbol{r}$ being the position of the particle, $q=\left\lvert \boldsymbol{r}_{ij}\right\rvert/h$, and $W_0=3/(359\pi h^3)$ for 3D problem.

At the right hand side of the discrete momentum equation, i.e. the first equation in \eqref{NSdiscrete}, the first term is the pressure gradient term using the TIC technique \cite{sun2018multi}, where
\begin{equation*}
		F_{ij}=-\left(p_i+p_j\right)+k_i,
\end{equation*}
\begin{equation*}
		k_i = \left\{
		\begin{aligned}
			&0,&&p_i \geqslant 0\text{ or }i \in D_f\\
			&2p_i,&&p_i <0\text{ and } i \notin D_f
		\end{aligned}
		\right.
		.
\end{equation*}
And $D_f$ is the particle set containing free-surface and its neighboring particles.
The second term is the viscous term and $K= 2(n+2)$ where $n$ is the number of spacial dimensions of the problem. $\pi_{ij}$ is the viscus interaction between particle $i$ and particle $j$,
\begin{equation*}
	\pi_{ij}=\frac{\left(\boldsymbol{u}_j-\boldsymbol{u}_i\right)\cdot\left(\boldsymbol{r}_j-\boldsymbol{r}_i\right)}{\left|\boldsymbol{r}_j-\boldsymbol{r}_j\right|^2} .
\end{equation*}
In the discrete continuity equation, i.e. the second equation in \eqref{NSdiscrete}, the second term at the right hand side is the artificial diffusion term \cite{marrone2011delta} where $\delta = 0.1$, and 
\begin{equation*}
	D_{ij} = 2\left[\left(\rho_j-\rho_i\right)-\frac{1}{2}\left(\left\langle \nabla\rho \right\rangle _i^L+ \left\langle \nabla\rho\right\rangle _j^L\right)\cdot\boldsymbol{r}_{ij}\right]\frac{\boldsymbol{r}_{ij}}{\boldsymbol{r}_{ij}^2} .
\end{equation*}
And $\left\langle \nabla \rho\right\rangle^L $ denotes the renormalized spatial gradient of $\rho$, \cite{antuono2010free}.

The pressure is straightforwardly linked to the density by the equation of state,
\begin{equation*}
	p=\frac{c_0^2\rho_0}{\gamma}\left(\left(\frac{\rho}{\rho_0}\right)^\gamma-1\right) ,
\end{equation*}
where $\rho_0$ is the initial density of fluid, $c_0$ is the numerical sound speed, $\gamma$ is usually set to be 7. To enforce the weakly compressible regime, $c_0$ is usually set to be larger than ten times of the maximum velocity in the computational domain.

To obtain a uniform particle distribution, the particle-shifting technique \cite{sun2017deltaplus} is also used. Positions of particles will be corrected at the end of each time step by
\begin{equation*}\label{shifting}
	\left\{ 
		\begin{aligned}
			&\boldsymbol{r}_i^* = \boldsymbol{r}_i + \delta \boldsymbol{r}_i \\
			&\delta\boldsymbol{r}_i = -C_{\mathrm{CFL}}\cdot \mathrm{Ma} \cdot \left(2h_i\right)^2 \cdot\sum_{i}\left[1+R\left(\frac{W_{ij}}{W\left(\Delta x\right)}\right)^{n'}\right]\nabla_i W_{ij} \frac{m_j}{\rho_i+\rho_j}
		\end{aligned}
	\right. ,
\end{equation*}
where $\boldsymbol{r}^*$ is the corrected particle position, $C_{\mathrm{CFL}}=1.5$, $R=0.2$, $n'=4$ and $\mathrm{Ma}=U_{0}/c_0$. $U_0$ is the reference velocity; in this work, it is set to be the inflow velocity or the structure velocity.

\section{Immersed boundary method for SPH}\label{IBM}
\begin{figure}[H]
	\centering
	\includegraphics[scale=0.5]{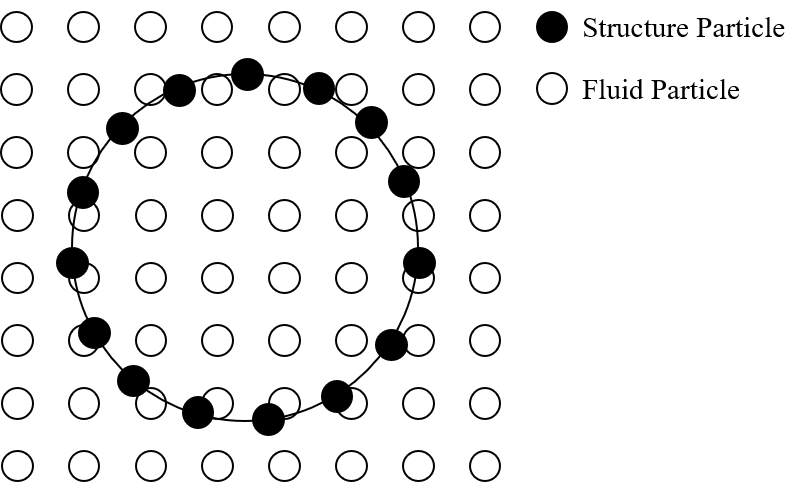}
	\caption{Immersed structure in fluid domain}
	\label{ibm}
\end{figure}

The IBM characters immersed structure boundaries by structural particles. We will see that these immersed structure particles, as illustrate in Figure~\ref{ibm}, play distinct roles compared to fluid particles.
Throughout this section, physical quantities associated with structural particles are denoted by uppercase symbols, e.g., $\boldsymbol{U}, \boldsymbol{R}, \boldsymbol{F}$, while those associated with fluid particles are denoted by lowercase symbols, e.g., $\boldsymbol{u}, \boldsymbol{r}, \boldsymbol{f}$. 

\subsection{MLS diffusive direct forcing IBM} \label{MLSIBM}
In the diffusive direct forcing scheme \cite{UHLMANN2005448}, the interaction between the structure boundaries and the fluid is computed through the motion of structure particles. At each time step $s_n$ from time instant $t_n$ to time instant $t_{n+1}$, $n\in\left\lbrace0, 1, 2,\cdots\right\rbrace$, the computation is divided into following three steps. Suppose the solution at time instant $t_{n}$ is known.

\paragraph{Step 1}
Velocity, position, and density of each fluid particle at time instant $t_{n+\frac{1}{2}}$ are calculated by
\begin{equation}\label{predictor}
	\begin{aligned}
		&\boldsymbol{u}_i^{n+\frac{1}{2}}=\boldsymbol{u}_i^n+ \frac{{\rm d}\boldsymbol{u}_i^n}{{\rm d}t} \frac{\Delta t}{2},\\
		&\boldsymbol{r}_i^{n+\frac{1}{2}}=\boldsymbol{r}_i^n+ \boldsymbol{u}_i^{n+\frac{1}{2}} \frac{\Delta t}{2},\\
		&\rho_i^{n+\frac{1}{2}}=\rho_i^n+\frac{{\rm d}\rho^n_i}{{\rm d}t}\frac{\Delta t}{2},
	\end{aligned}
\end{equation}
where $\Delta t$ is the time interval, i.e. $\Delta t = t_{n+1}-t_{n}$.
The desired velocity and position of the structure particle at time instant $t_{n+1}$ are calculated,
\begin{equation}\label{desiredU}
	\begin{aligned}
		&\boldsymbol{U}^d_b=\boldsymbol{U}^n_b+\boldsymbol{\rm RHS}^{n+\frac{1}{2}}_b \Delta t,\\
		&\boldsymbol{R}^d_b=\boldsymbol{R}^n_b + \boldsymbol{U}^n_b \Delta t,
	\end{aligned}
\end{equation}
where $\boldsymbol{\rm RHS}_b$ represents the acceleration of structure particle indexed $b$ which is typically given by the problem setup. \eqref{desiredU} provides the calculation method for moving structures, and in this study, we only consider the case of uniform-speed structures, i.e., $\boldsymbol{\rm RHS}_b=\boldsymbol{0}$.

\paragraph{Step 2}
The preliminary velocity field at time instant $t_{n+1}$ is evaluated by
\begin{equation}\label{preliminart_velocity}
	\boldsymbol{u}^*_i = \boldsymbol{u}^n_i + \frac{{\rm d}\boldsymbol{u}^{n+{\frac{1}{2}}}_i}{{\rm d}t}\Delta t .
\end{equation}
Using the kernel integration, the preliminary velocity at the position of structure particle $b$ is
\begin{equation*}\label{Velocity}
	\boldsymbol{U}^*_b = \sum_{j}^{N_f} \boldsymbol{u}_j^* W_{bj} V_j,
\end{equation*}
where $N_f$ is the total number of fluid particles within the supporting dimain. 

For no-slip boundaries, the velocity of fluid particle at the position of structure particle is equal to that of the structure particle. This means the boundary force at the position of structure particle indexed $b$ must force the fluid velocity at time instant $t_{n+1}$ to be $\boldsymbol{U}_b^d$. Therefore, the force can be computed,
\begin{equation*}\label{Force}
	\boldsymbol{F}_b^{n+\frac{1}{2}}=\rho_0\frac{\boldsymbol{U}_b^d-\boldsymbol{U}_b^*}{\Delta t}.
\end{equation*} 

In the diffusive direct forcing scheme, the forces on fluid particles near the structure are obtained based on the kernel interpolation, i.e.,
\begin{equation}\label{force}
	\boldsymbol{f}_i^{n+\frac{1}{2}} = \sum_{b}^{N_b}\boldsymbol{F}_b^{n+\frac{1}{2}} W_{ib} V_b,
\end{equation}
where $N_b$ and $V_b$ are the total number of structure particles within supporting domain and the volume of the structure particle indexed $b$, respectively. \eqref{force} can be interpreted as the propagation of structure-fluid interation. It can be proven that momentum conservation is equivalent to the normalization condition of kernel function \cite{NASAR2019263},
\begin{equation} \label{conservation}
	\sum_{b}^{N_b} W_{ib} V_b  = 1.
\end{equation}

Using the IBM, the structure boundary only consist of one layer of structure particles, thus \eqref{conservation} can not be fully satisfied. To address this limitation, Nasar et al. correct the shape of the kernel function based on MLS method. The 2D MLS kernel function is defined as
\begin{equation} \label{Wmls}
	W_{ib}^{\mathrm{MLS}}=\left(\beta_i\cdot\left(1,x_{ib},y_{ib}\right)\right)W_{ib},
\end{equation}
where $x_{ib} = x_i - x_b$, $y_{ib} = y_i - y_b$, $\beta_i=A_i^{-1}[\begin{matrix}1 & 0 & 0\end{matrix}]^{\rm T}$
and 
\begin{equation*} \label{Amls}
	A_i = \sum_{j=1}^{N}\left[
	\begin{matrix}
		1          &x_{ib}       &y_{ib}\\
		x_{ib}    &x_{ib}^2     &x_{ib}y_{ib}\\
		y_{ib}    &x_{ib}y_{ib}  &y_{ib}^2\\
	\end{matrix}
	\right]W_{ib}V_b.
\end{equation*}
An introduction of the MLS method is presented in \ref{appendix}, including the kernel function in 3D.


\paragraph{Step 3}
Based on the external force term computed in \eqref{force}, the expression for the physical quantities at time instant $t_{n+1}$ is
\begin{equation}\label{corrector}
	\begin{aligned}
		&\boldsymbol{u}_i^{n+1}=\boldsymbol{u}_i^n+\frac{{\rm d}\boldsymbol{u}_i^{n+\frac{1}{2}}}{{\rm d}t}\Delta t + \frac{\boldsymbol{f}_i^{n+\frac{1}{2}}}{\rho_i} \Delta t,\\
		&\boldsymbol{r}_i^{n+1}=\boldsymbol{r}_i^n+ \boldsymbol{u}_i^{n+\frac{1}{2}}\Delta t, \\
		&\rho_i^{n+1}=\rho_i^n+\frac{{\rm d}\rho^{n+\frac{1}{2}}_i}{{\rm d}t}\Delta t.
	\end{aligned}
\end{equation}
The position and velocity of structure particles are updated by
\begin{equation}\label{update_structure}
	\begin{aligned}
		\boldsymbol{R}_i^{n+1} = \boldsymbol{R}_i^d ,\\
		\boldsymbol{U}_i^{n+1} = \boldsymbol{U}_i^d	.
	\end{aligned}
\end{equation}

\subsection{MLS direct forcing IBM}
One of the fundamental limitations in the MLS diffusive direct forcing scheme is the potential instability in 3D simulations. This instability arises from an insufficient number of interpolation points and results in singular matrices $A$. To address this issue, we employ the direct forcing scheme for the force computation. Different from \eqref{force}, the force calculation in the proposed method is defined as
\begin{equation}\label{directforcing}
	\boldsymbol{f}_i^{n+\frac{1}{2}}=\rho_0\frac{\boldsymbol{u}_i^{d} - \boldsymbol{u}_i^*}{\Delta t},
\end{equation} 
where $\boldsymbol{u}^{d}$ is the desired velocity of the fluid at time instant $t_{n+1}$. In the direct forcing scheme, the desired velocity of particles far from the boundary, comes from the solution of momentum equation \eqref{NS} without infulence from the boundary; while that of the position at structure particles equals to the structure velocity. The desired velocity of fluid near the structure is determine by a interpolation from that two kinds of particles. As mentioned in \cite{UHLMANN2005448}, the linear interpolation may lead to strong oscillations due to insufficient smoothing. In the SPH, it becomes even more formidable due to the ununiform distribution of particles. As an interpolation method, the MLS exhibits a superior smoothness in handling data with irreglar distributions. In this works, it is used to determine the desired velocity near the srtucture.


\begin{figure}[H]
	\centering
	\subfloat[]{ \label{extrapolation}
		\includegraphics[scale=0.5]{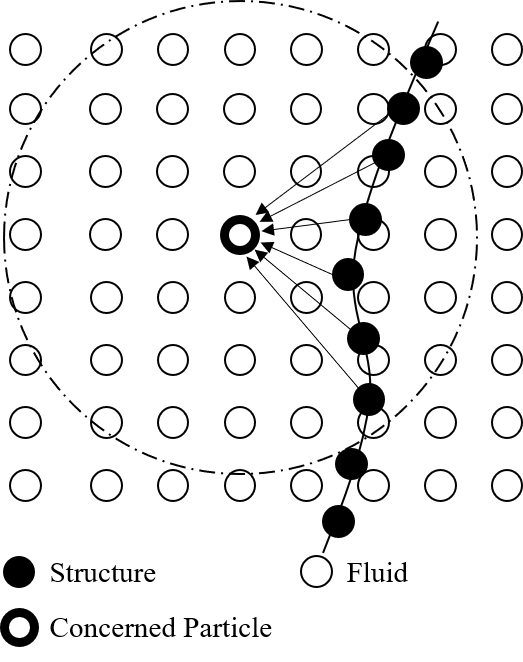}
	}
	\hspace{0.5in}
	\subfloat[]{ \label{interpolation}
		\includegraphics[scale=0.5]{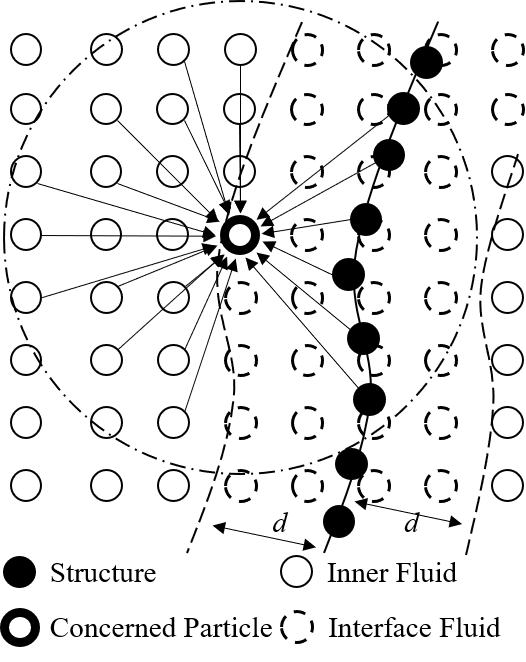}
	}
	\caption{Schematic of two interpolation schemes: (a) Diffusive direct forcing scheme, (b) Direct forcing scheme}
\end{figure}

As shown in Figure~\ref{interpolation}, particles located within a distance $d$ from the structure surface are labeled as interface fluid particles, while all other particles are labeled as inner fluid particles. $\boldsymbol{u}^d$ and $\boldsymbol{f}$ are calculated only on the interface fluid particles. $\boldsymbol{u}^d$ of the interface fluid particle indexed $i$ is calculated as:
\begin{equation}\label{interpolation_velocity}
	\boldsymbol{u}^d_i=\sum_{j}^{N_{i}} \boldsymbol{u}^*_j W_{ij}^{\mathrm{MLS}} + \sum_{j}^{N_b} \boldsymbol{U}^d_j W_{ij}^{\mathrm{MLS}},
\end{equation}
where $N_{i}$ and $N_b$ are the numbers of the inner fluid particles and structure particles within the supporting domain. The distance $d$ usually set to be half radius of the supporting domain. This can reduce the influence from the velocity at other side of the structure as well as can maintain a smooth interpolation. In 3D problems, $W_{ij}^{\mathrm{MLS}}$ is
\begin{equation}\label{W3d}
	W_{ij}^{\mathrm{MLS}} = 
	\left[
	\begin{matrix}
		1 &0 &0 &0
	\end{matrix}
	\right]
	A_i^{-1}
	\left[
	\begin{matrix}
		1\\
		x_{ij}\\
		y_{ij}\\
		z_{ij}\\
	\end{matrix}	
	\right]	W_{ij},
\end{equation}
where $\left[ \begin{matrix}
	x_{ij} &y_{ij} &z_{ij} 
\end{matrix} \right]^T
= \left[
\begin{matrix}
	x_{i}-x_{j} &y_{i}-y_{j} &z_{i}-z_{j} 
\end{matrix} 
\right]^T$ is the distance between particles indexed $i$ and $j$, $A$ is evaluated based on the position of interface particle and structure surface point,
\begin{equation*}\label{A3d}
	A_i = \sum_j^{N_i+N_b}	\left[
	\begin{matrix}
		1	 &x_{ij}     &y_{ij}     &z_{ij}\\
		x_{ij}  &x_{ij}^2 &x_{ij} y_{ij} &x_{ij} z_{ij}\\
		y_{ij}  &y_{ij} x_{ij} &y_{ij}^2 &y_{ij} z_{ij} \\
		z_{ij}   &z_{ij} x_{ij} &z_{ij} y_{ij} &z_{ij}^2 \\
	\end{matrix}	
	\right] W_{ij}.
\end{equation*}

In summary, the computational process of the proposed method explained in this section is illustrated as follows.
\begin{algorithm}\label{time_advancing_step}
	\caption{Time advancing step}
	\KwIn{The velocit $\boldsymbol{u}^{n}$,$\boldsymbol{U}^{n}$, position $\boldsymbol{r}^{n}$,$\boldsymbol{R}^{n}$ and fluid density $\rho^n$ at time step $n$}
	\KwOut{The velocity $\boldsymbol{u}^{n+1}$,$\boldsymbol{U}^{n+1}$, position $\boldsymbol{r}^{n+1}$,$\boldsymbol{R}^{n+1}$ and fluid density $\rho^n$ at time step $n+1$}
	Searching the neighbor particles within supporting domain\; 
	
	Calculate
	$\dfrac{{\rm d}\boldsymbol{u}^{n}}{{\rm d}t}$,$\dfrac{{\rm d}\rho^{n}}{{\rm d}t}$ based on \eqref{NSdiscrete}\;
	
	Calculate $\boldsymbol{u}^{n+1/2}$, $\boldsymbol{r}^{n+1/2}$ and $\rho^{n+1/2}$ based on \eqref{predictor}\;
	
	Calculate the desired velocity $\boldsymbol{U}^d$ and the desired position $\boldsymbol{R}^d$ based on \eqref{desiredU}\;	
	
	Update the neighbor particles list;
	
	Calculate
	$\dfrac{{\rm d}\boldsymbol{u}^{n+1/2}}{{\rm d}t}$,$\dfrac{{\rm d}\rho^{n+1/2}}{{\rm d}t}$ based on \eqref{NSdiscrete}\;

	Predict the preliminart velocity
	$\boldsymbol{u}^*$ at time $n+1$ based on \eqref{preliminart_velocity}\;
	
	Determine the interface particles and inner particles as Figure.~\ref{interpolation}\;
	
	Interpolate the desired velocity on fluid particle $\boldsymbol{u}^d$ based on \eqref{interpolation_velocity}\;
	
	Calculate the force $\boldsymbol{f}^{n+1/2}$ based on \eqref{directforcing}\;
	
	Correct the velocity $\boldsymbol{u}^{n+1}$, position $\boldsymbol{r}^{n +1}$ and density $\rho^{n+1}$ of fluid particles based on \eqref{corrector}\;
	
	Update the velocity $\boldsymbol{U}^{n+1}$ and position $\boldsymbol{R}^{n +1}$ of strcture particles based on \eqref{update_structure}
\end{algorithm}


\section{Numerical tests}\label{result}
In this section, three numerical tests are presented to illustrate the performance of the proposed method. The first one is the impulsively started plate test at Reynolds number $\mathrm{Re}=126$, and the other two simulate the flow past a infinite cylinder pipe at $\mathrm{Re}=20$ and $\mathrm{Re}=200$, respectively.

\subsection{Impulsively started plate test} \label{plate}
The impulsively started plate test is a classic fluid dynamics experiment used to study the response of fluid to a sudden application of an external force. In numerical studies, it is usually simplified to a one-way coupled problem. The immersed thin plate is impulsively started and moves through the fluid at a constant velocity. After the impulsive start, the fluid exhibits an inertial response, and vortices are formed on the surface of the plate. It is worthy to evaluate the stability and accuracy of the proposed IBM scheme through this test case.

\begin{figure}[h!]
	\centering
	\includegraphics[scale=0.6]{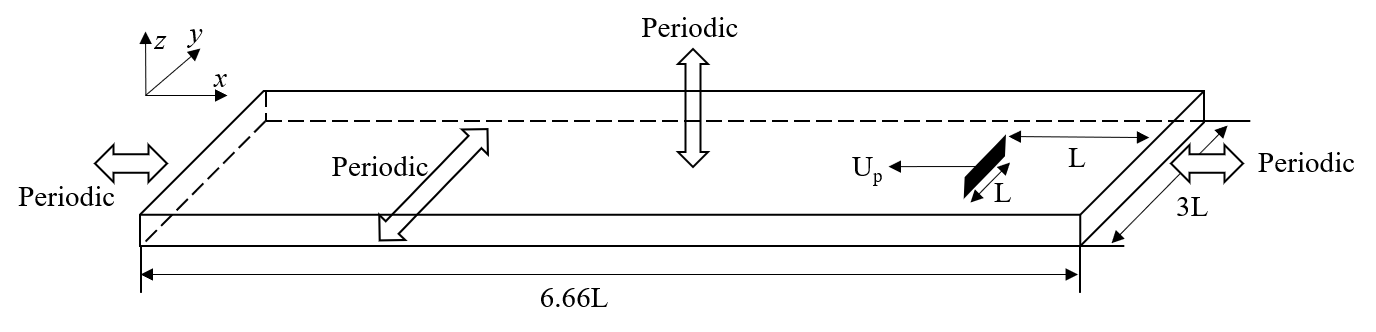}
	\caption{Computation domain for a 3-D thin rigid impulsively started plate}	
	\label{platemodel}
\end{figure}

The configuration of this test is shown in Figure~\ref{platemodel}. The black plate extends infinitely along the $z$-direction with a width of $L=0.3m$. The fluid is static initially. A constant velocity $U_p=0.1m/s$ along the direction indicated by the arrow is applied to this plate. The plate is represented by a series IBM particles. The simulation is conducted at 
\begin{equation*}
	\mathrm{Re}=\frac{LU_p\rho}{\mu}=126.
\end{equation*}
The size of the computational domain along the $x$-direction and $y$-direction is $6.66L$ and $3L$, respectively. And the height of the domain is $5\Delta x$. The domain is periodic in all directions. In this study, we use $\Delta x = L/120$, $\Delta t = 3.5\times10^{-5}s$ and $c_0 = 15m/s$. This configuration is same to that of \cite{NASAR2019263}. We will use $\boldsymbol{u}$ to denote the velocity of the flow relative to the plate. And non-dimensional time
\begin{equation*}
	t^* = tU_p/L
\end{equation*}
is also utilized.

\begin{figure}[H]
	\centering
	\subfloat[$ \frac{p}{\rho \boldsymbol{U}_p^2}$]{
		\includegraphics[scale=0.3,trim=0 0 0 65, clip]{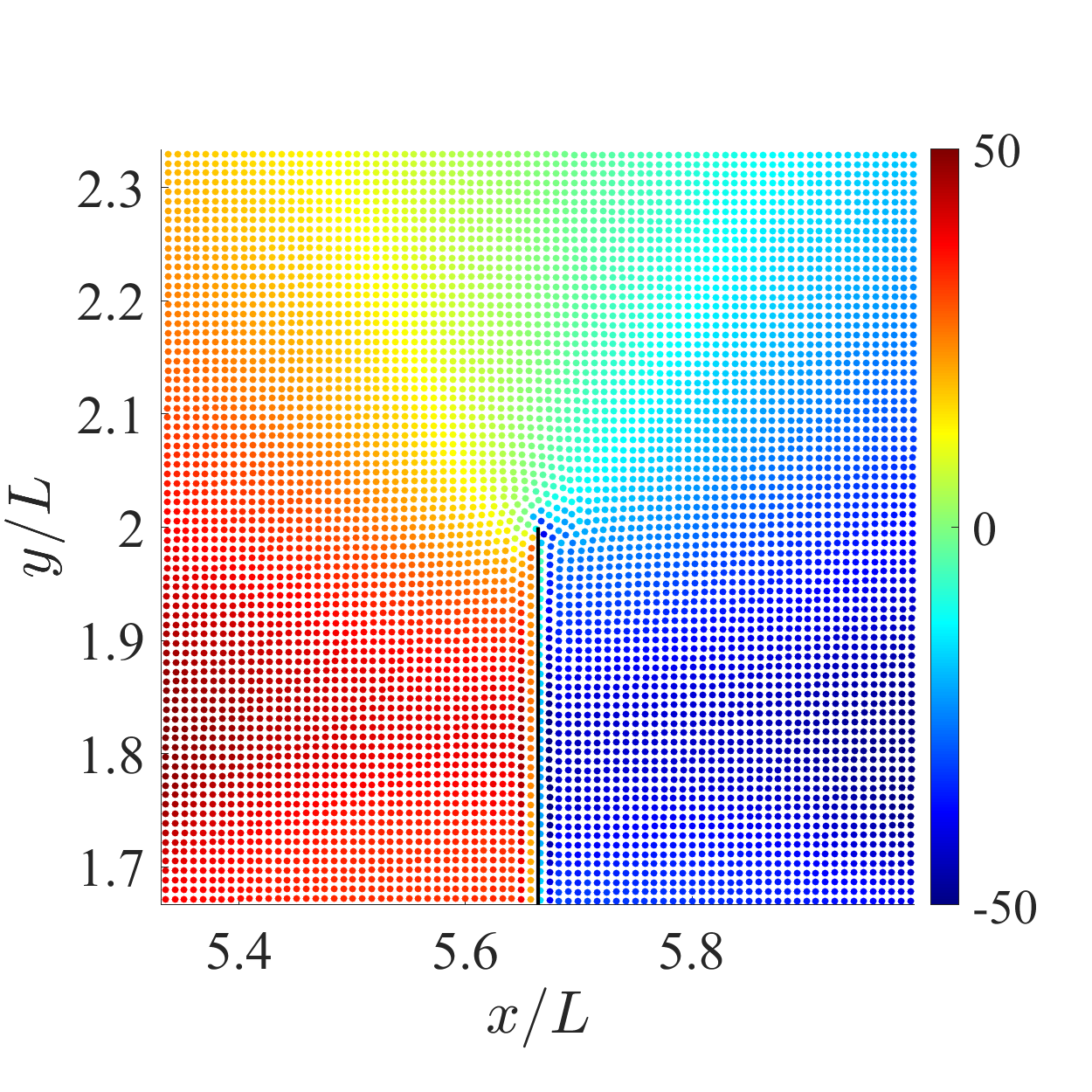}
		\label{500pre}
	}
	\subfloat[$\left\| \boldsymbol{u}\right\| $]{
		\includegraphics[scale=0.3,trim=0 0 0 65, clip]{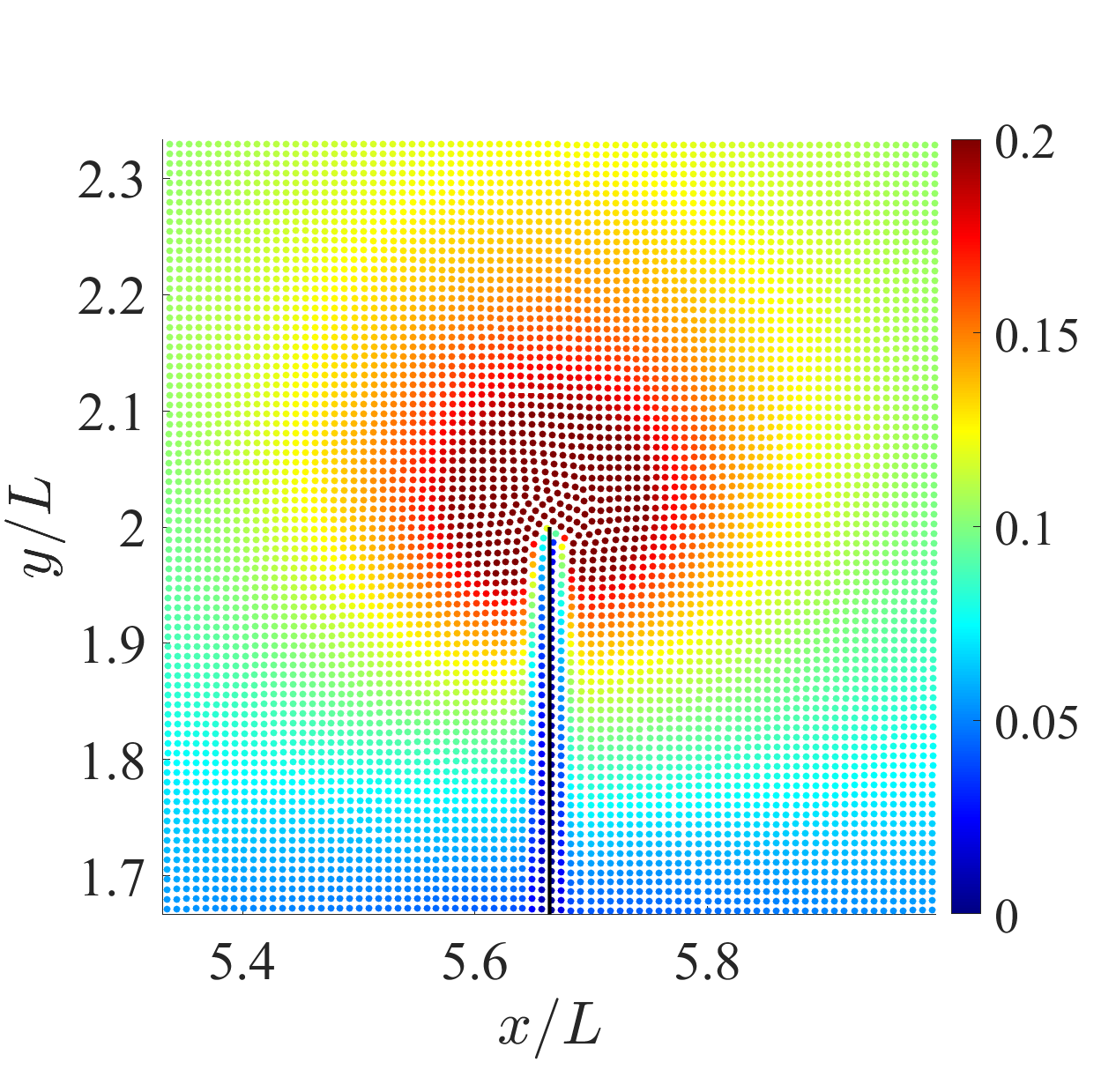}
		\label{500vel}
	}
	\caption{pressure and velocity at the plate edge, $t^*$ is $0.0058333$}
	\label{preandvel}
\end{figure}

\begin{figure}[H]
	\centering
	\subfloat{
		\includegraphics[scale=0.3,trim=0 20 0 20, clip]{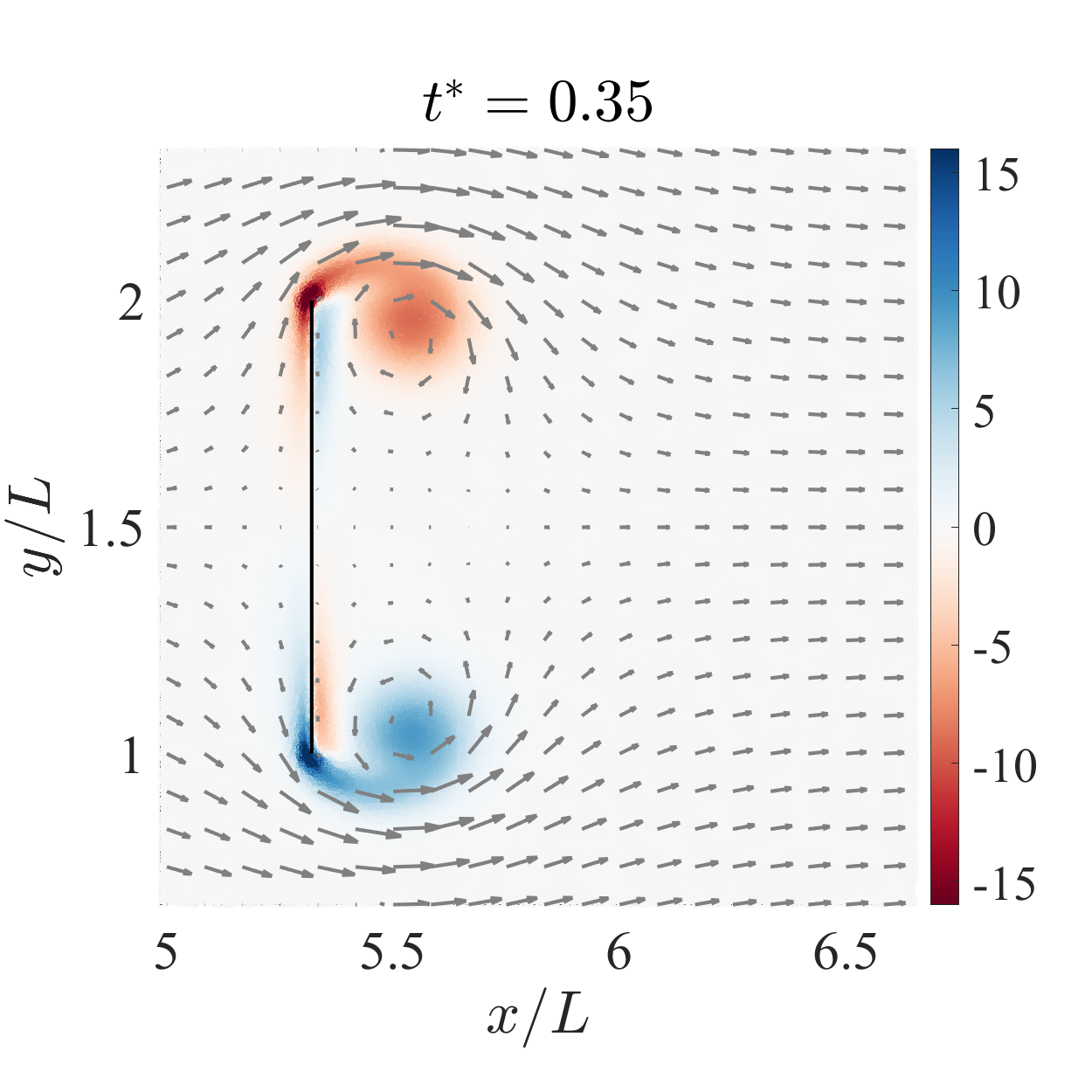}
	}
	\subfloat{
		\includegraphics[scale=0.3,trim=0 20 0 20, clip]{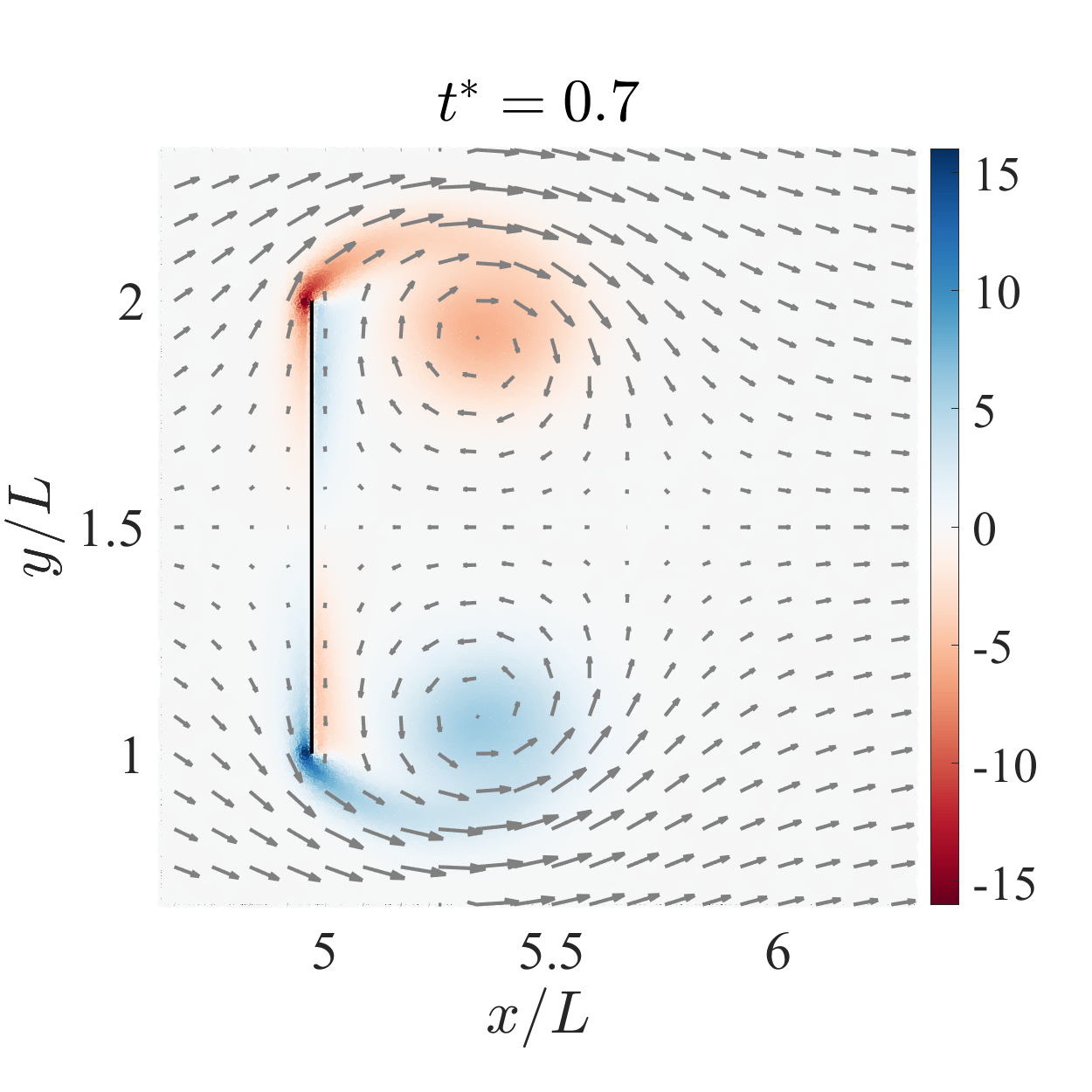}
	}\\
	\subfloat{
		\includegraphics[scale=0.3,trim=0 20 0 20, clip]{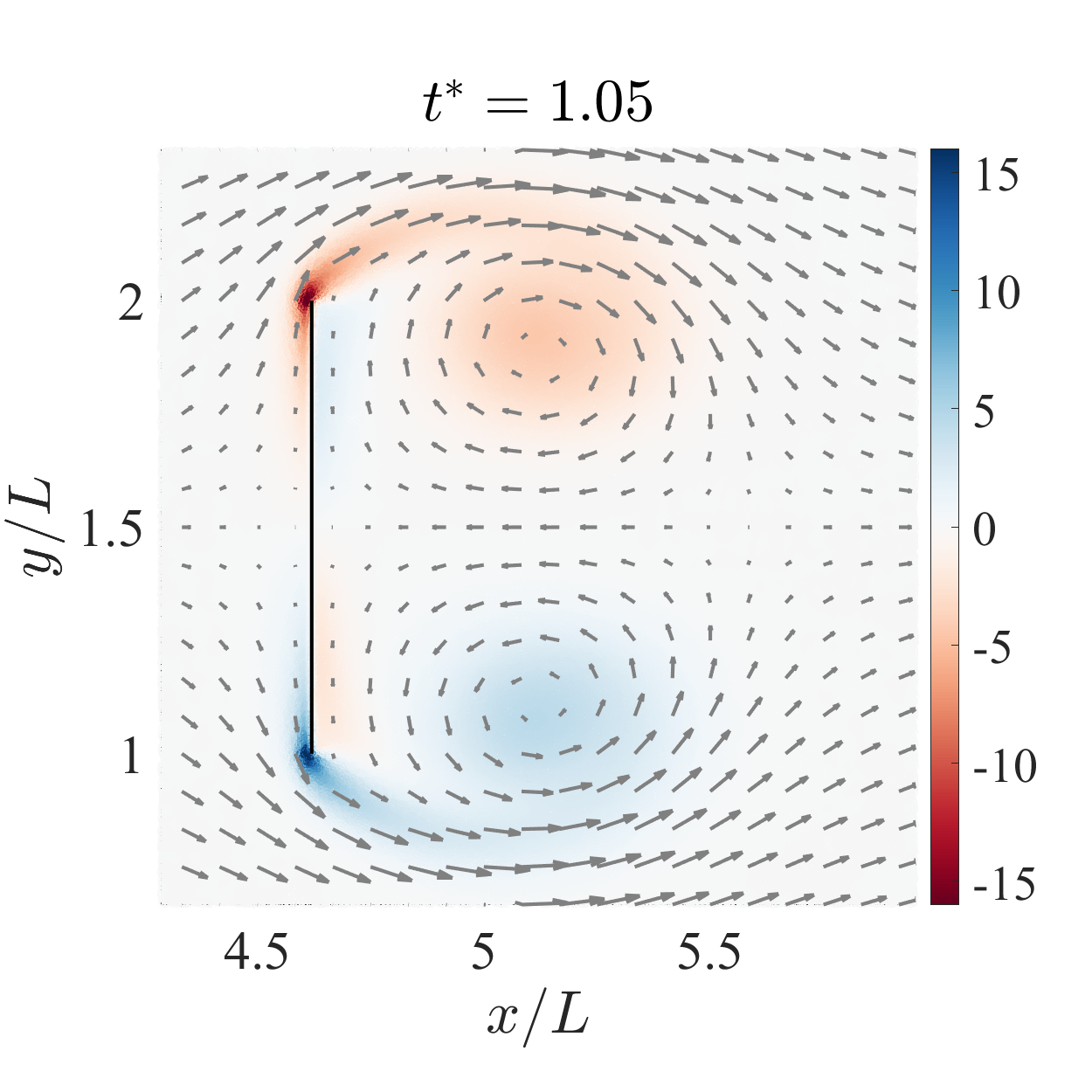}
	}
	\subfloat{
		\includegraphics[scale=0.3,trim=0 20 0 20, clip]{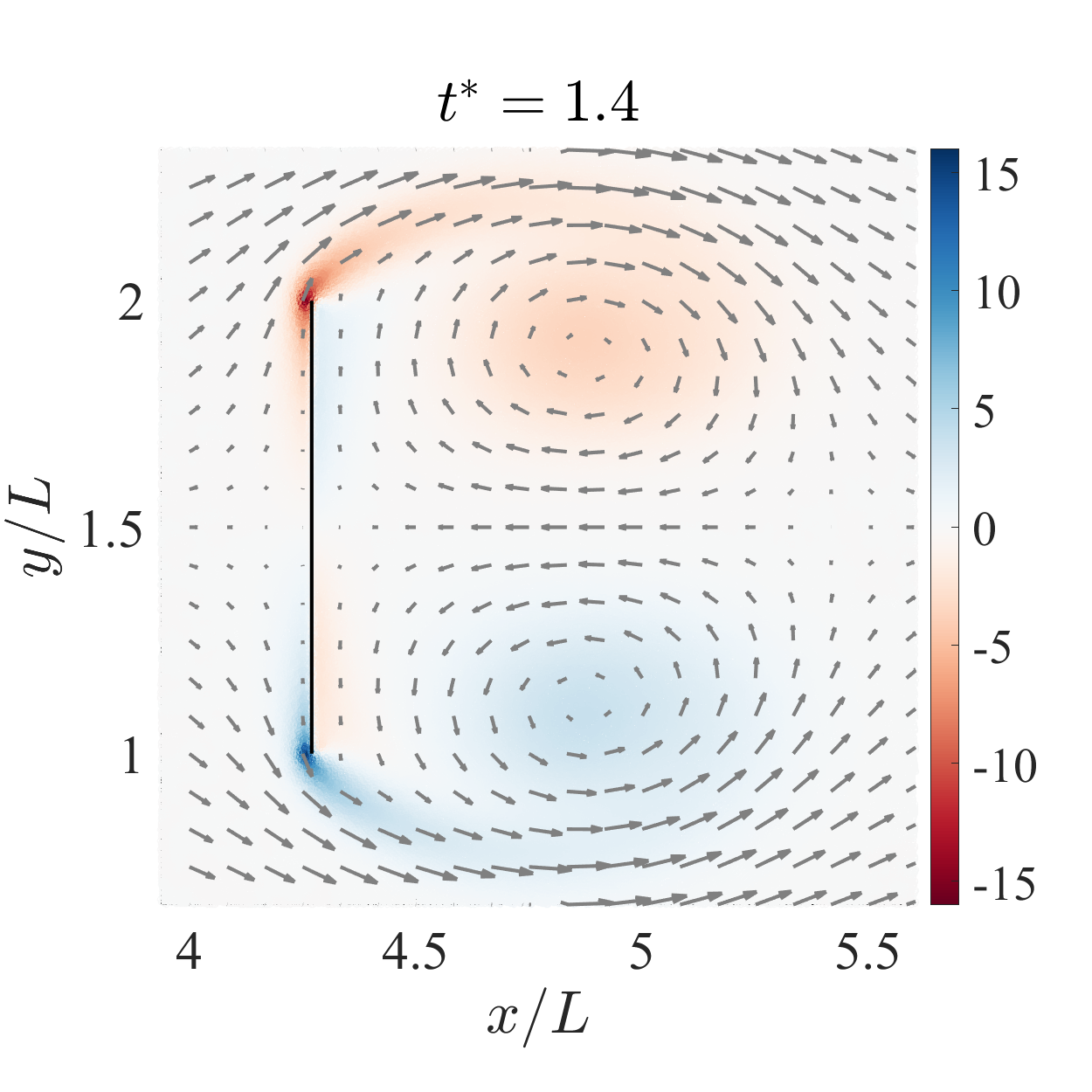}
	}
	\caption{Vorticity in $z$-direction and velocity vector in the $xy$-plane}
	\label{vortexbubble}
\end{figure}

At the first stage of the simulation, there exists a velocity discontinuity of the plate and the fluid. It can lead to instability in the numerical results. Thus the smoothness in the interpolation process is required. Figure~\ref{preandvel} illustrates the pressure and velocity fields near the plate edge at $t^*=0.0058333$. As shown in Figure~\ref{500pre}, the impulsive start of the plate induces a significant pressure difference between the two sides of the plate. And as $y$ increases, the pressure gradually diffuses, exhibiting a smooth transition. Figure~\ref{500vel} presents the distribution of the vector norm of the velocity relative to the plate, i.e. $\left\lVert \boldsymbol{u}\right\rVert := \sqrt{\boldsymbol{u} \cdot \boldsymbol{u}}$. It is close to zero near the plate surface. And it reaches approximately $0.1m/s$ far away from the plate. Near the plate edge, there is a high velocity region of vector norm around $0.2m/s$. The transition between these velocity regions is smooth. Thus, in this case, the MLS velocity interpolation method used in our study achieves smooth interpolation of the velocity field, ensuring computational stability.

\begin{figure}[H]
	\centering
	\includegraphics[scale=0.3]{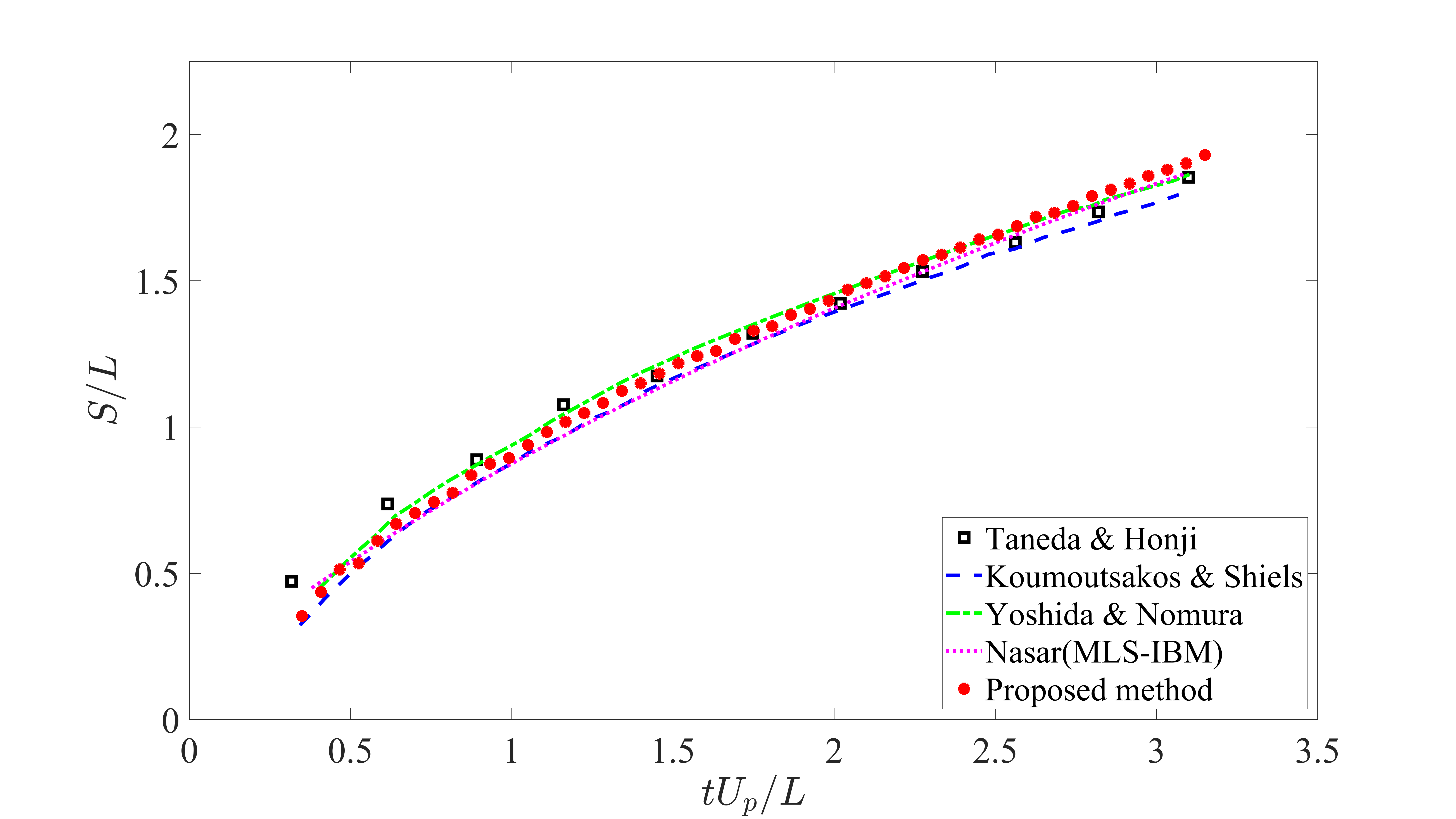}
	\caption{Movement of the stagnation point}
	\label{length}
\end{figure}

Figure~\ref{vortexbubble} shows the development of the vortex bubbles induced by the suddenly moving plate and the velocity vector at time $t^*=0.35, 0.7, 1.05$ and $1.4$. The velocity vector is presented on a grid of interval $\Delta x_{\mathrm{grid}}=10\Delta x$. It is seen that vortices appear at both edges of the plate and move away from the plate edges. Their shapes develop from circles to ellipses and the vorticity gradually decrease. These vortices exert an influence extending from the plate to a distance behind it. Within this span, there is a reversal in fluid velocity, transitioning from the initial flow direction to its opposite. As the distance progresses, the fluid gradually returns to its normal flow direction. The stagnation point where the velocity starts to reverse is commonly used to measure the length of the wake vortices. Figure~\ref{length} shows results of the distance between the stagnation point and the plate. A good match to the references taken from \cite{taneda1971unsteady, koumoutsakos1996simulations, yoshida1985transient, NASAR2019263}, is observed.

\subsection{Flow past an infinitely long cylinder pipe}\label{cylinder}

\begin{figure}[H]
	\centering
	\includegraphics[scale=0.6]{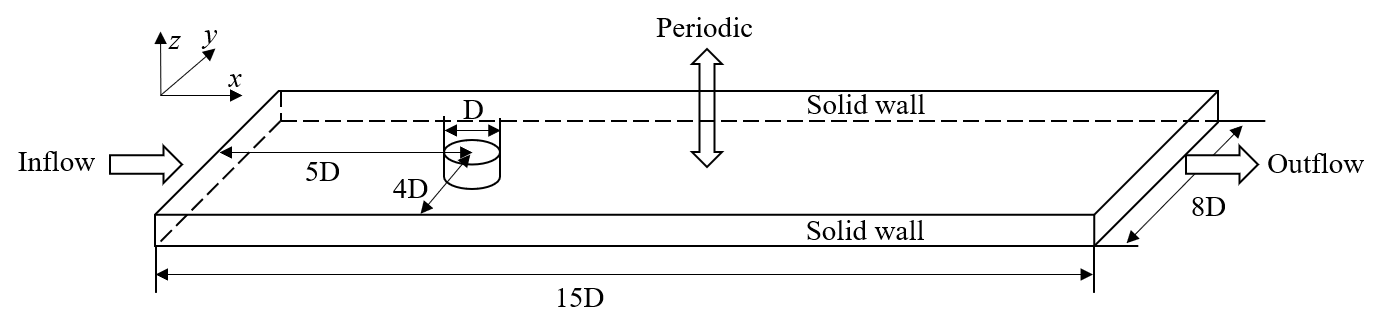}
	\caption{Computation domain for 3-D flow past a infinitely long cylinder}
	\label{cylindermodel}
\end{figure}

The configuration of this test case is illustrated in Figure~\ref{cylindermodel}. The model consists of an infinitely long cylindrical pipe with a diameter of $D=0.1m$. The size of the computation domain in $x$ and $y$-directions are $15D$ and $8D$ respectively. The domain is periodic in $z$-direction. The $x^-$-face is the fluid inlet implemented using the mirror method, and the $x^+$-face is the fluid outlet implemented with the do-nothing method. Same configurations for inlet and outlet conditions can be found in Pawan's work \cite{negi2020improved}. To prevent the instability induced by repulsively starting, the velocity field starts with a resting state, and gradually increases during $t\in\left[0,D/U_{\mathrm{in}}\right]$, where $U_{\mathrm{in}}=1m/s$ is the final velocity of inflow; a constant acceleration $a_0=U_{\mathrm{in}}^2/D$ is applied during this period.

Note that, in this case, only the cylinder surface is modeled using the IBM. The boundaries on both sides of $y$-direction are modeled as solid boundaries same as $\delta$-SPH \cite{marrone2012study}. The mirror point interpolation is used to calculate the pressure of these solid wall particles. The velocity of wall particles are zero. To reduce the influence to the fluid field, these particles do not participate the calculation of the viscus term in \eqref{NSdiscrete}. In this case, $\Delta x=D/30$, $\Delta t = 1\times 10^{-4}$, which leads to stable results, see Thanh's work \cite{bui2021simplified}. The numerical speed of sound is set to $c_0=20m/s$. Reynolds number is calculated by
\begin{equation*}
	\mathrm{Re}=\rho U_{\mathrm{in}}D/\mu.
\end{equation*}
And we use dimensionless time
\begin{equation*}
	t^* = tU_{\mathrm{in}}/D.
\end{equation*}

\begin{figure}[H]
	\centering
	\subfloat[Diffusive direct forcing method]{ \label{vort96re20dif}
		\includegraphics[scale=0.25,trim=130 0 140 0, clip]{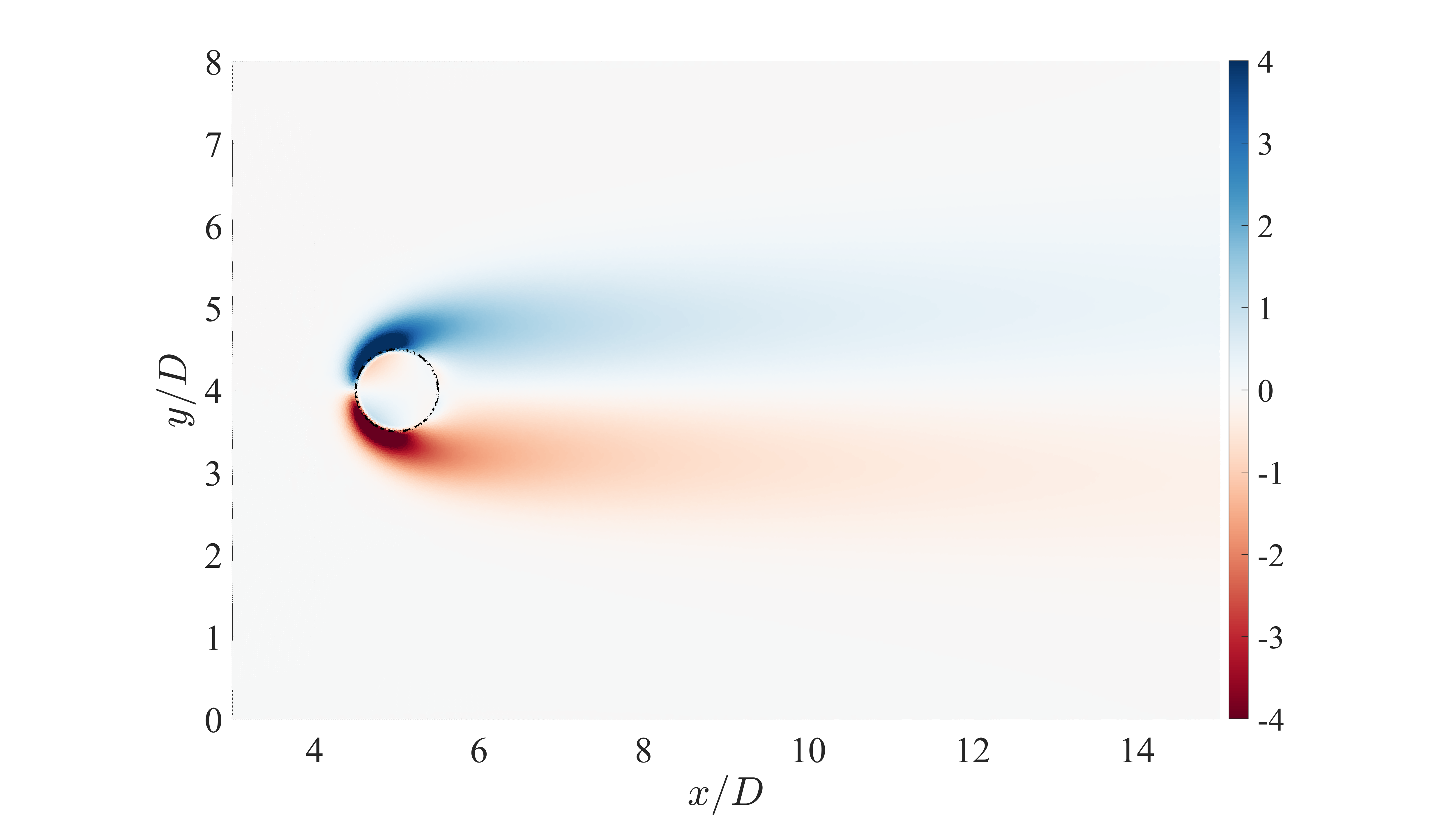}
	}
	\subfloat[Proposed method]{ \label{vort96re20mls}
		\includegraphics[scale=0.25,trim=130 0 140 0, clip]{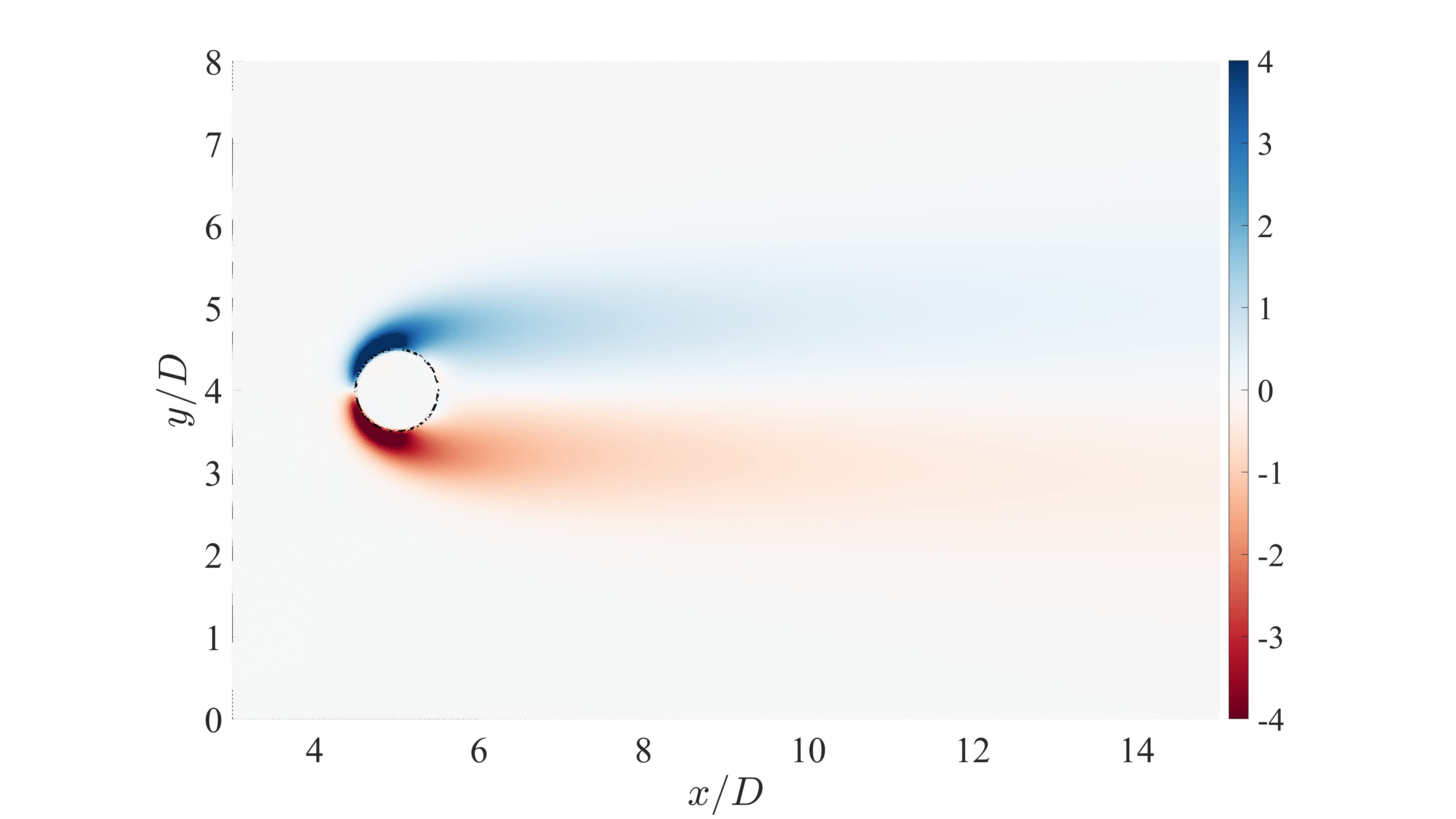}
	}
	\caption{Vorticity field at $t^* = 96$ and $\mathrm{Re} = 20$}
	\label{vort96re20}
\end{figure}

\begin{figure}[H]
	\centering
	\subfloat[Diffusive direct forcing method]{ \label{t96re200dif}
		\includegraphics[scale=0.25,trim=130 0 140 0, clip]{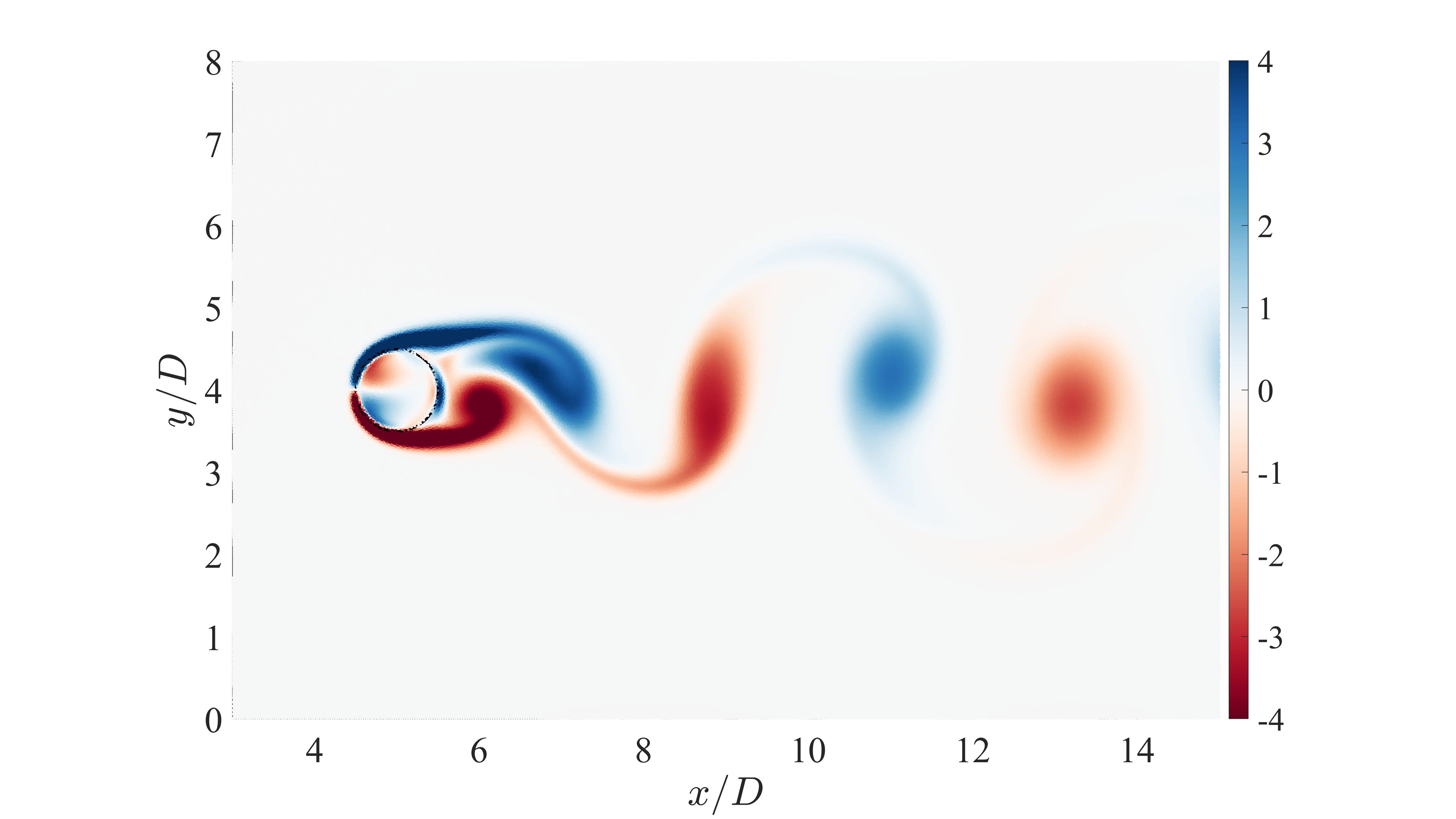}
	}
	\subfloat[Proposed method]{ \label{t96re200mls}
		\includegraphics[scale=0.25,trim=130 0 140 0, clip]{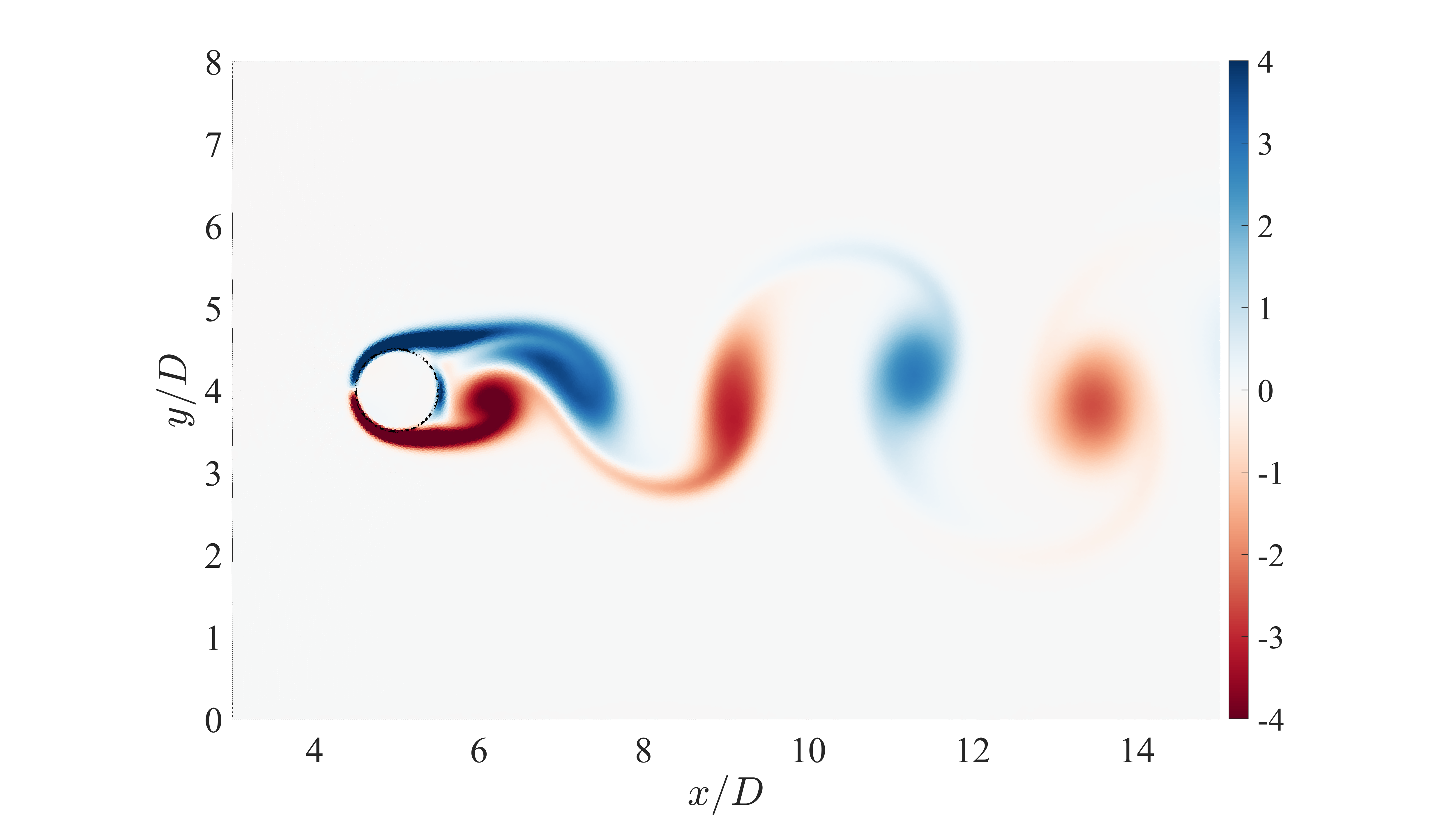}
	}
	\caption{Vorticity field at $t^* = 96$ and $\mathrm{Re} = 200$}
	\label{vort96re200}
\end{figure}

Figure~\ref{vort96re20} and Figure~\ref{vort96re200} show results of the $z$-component of the dimensionless vorticity
\begin{equation*}
	\omega_z^*=\omega_z D/U_{\mathrm{in}},
\end{equation*}
at time $t^*=96$ for $\mathrm{Re}=20$ and $\mathrm{Re}=200$ with both the diffusive direct forcing (hereafter referred to as DDF) method and the proposed method. (The DDF method is the non-MLS version of the approach discussed in Section~\ref{MLSIBM}. For the MLS DDF, the MLS matrix $A$ of some particles becomes singular in this case, which leads to a quick divergence.) At $\mathrm{Re}=20$, results of both method show stable wake patterns. Similarly, at $\mathrm{Re} = 200$, both methods show periodic vortex shedding. Inside the cylinder, the DDF method exhibits a non-zero vorticity field. This is because when spreading the force from structure particles to fluid particles, the force act on both sides. On the contrary, the proposed method effectively maintains a zero vorticity field within the cylinder.

\begin{figure}[H]
	\centering
	\includegraphics[scale=0.5, trim=0 0 0 0]{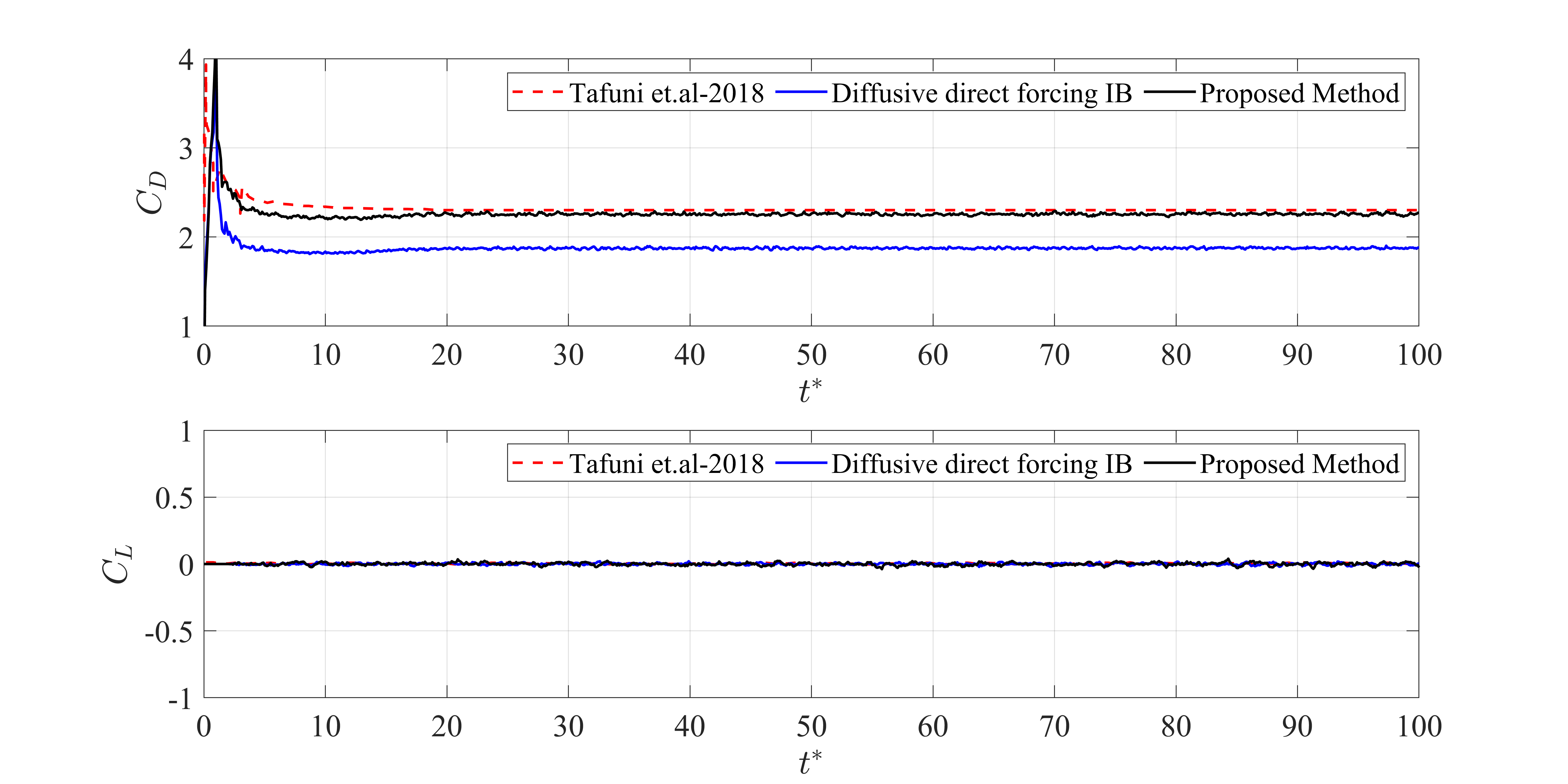}
	\caption{Time history of the drag and lift coefficients for flow past a circular cylinder at $\mathrm{Re} = 20$}
	\label{C_20}
\end{figure}

\begin{figure}[H]
	\centering
	\includegraphics[scale=0.5, trim=0 0 0 0]{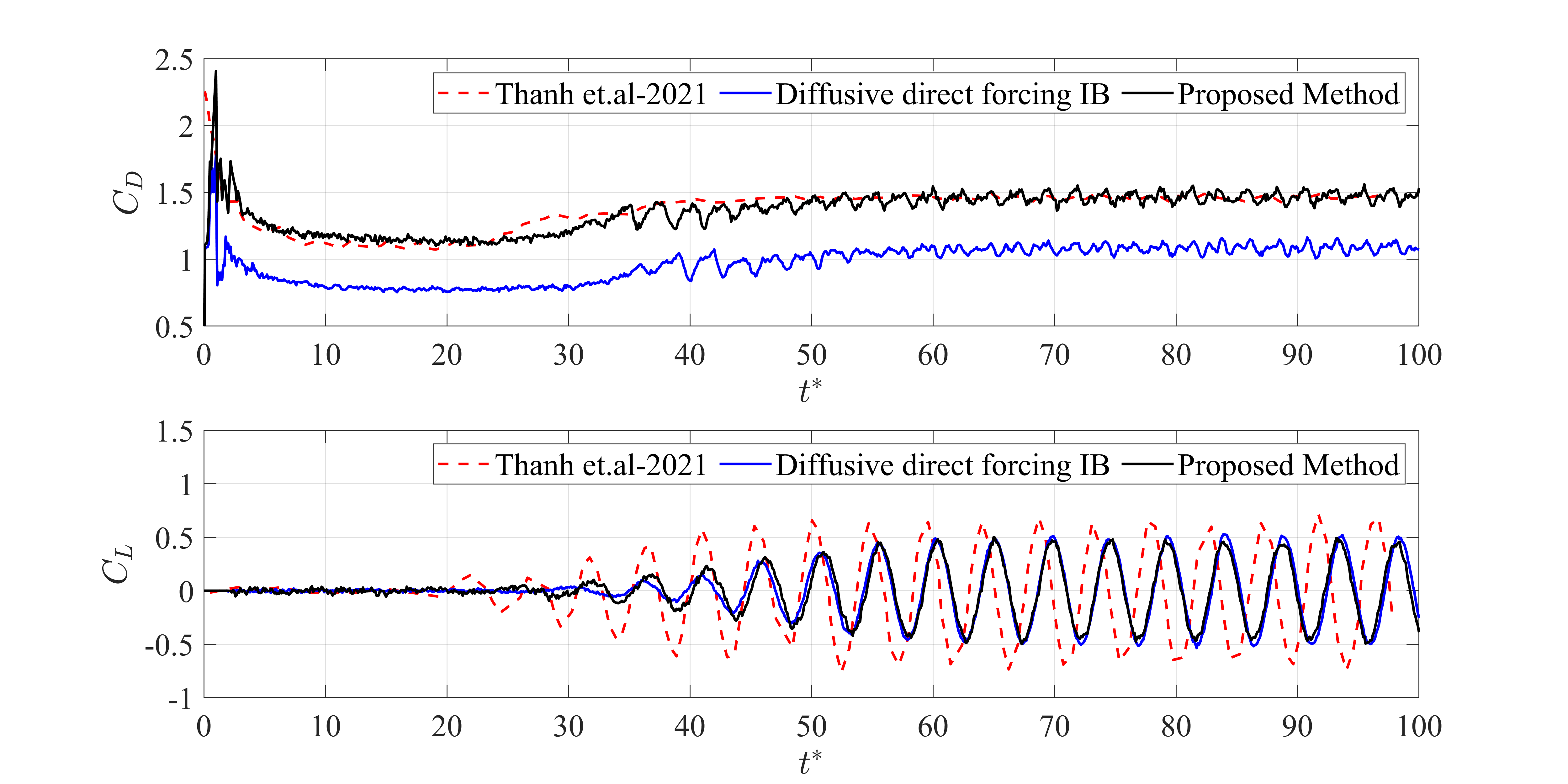}
	\caption{Time history of the drag and lift coefficients for flow past a circular cylinder at $\mathrm{Re} = 200$}
	\label{C_200}
\end{figure}

Figure~\ref{C_20} and Figure~\ref{C_200} present results of drag and lift coefficients,
\begin{equation*}
	C_D = \frac{2F_{\mathrm{total},x}}{\rho\boldsymbol{U}_{\mathrm{in}}^2 D} ,\quad C_L = \frac{2F_{\mathrm{total},y}}{\rho\boldsymbol{U}_{\mathrm{in}}^2 D},
\end{equation*}
 where $F_{\mathrm{total},x}$ and $F_{\mathrm{total},y}$ are the $x$ and $y$ components of the total force per unit length, i.e. $\boldsymbol{F}_{\mathrm{total}}$, on the cylinder surface. The total force is evaluated using a surface integral of the stress tensor over the cylinder,
\begin{equation*}\label{totalforce}
	\boldsymbol{F}_{\mathrm{total}} = \int_{\Omega}\left(
	-p\boldsymbol{n}+\mu\left(
	\nabla\cdot\boldsymbol{u}+\left(\nabla\cdot\boldsymbol{u}\right)^T
	\right)\cdot\boldsymbol{n}
	\right)dS ,
\end{equation*}
where $\Omega$ is the surface area per unit length of the cylinder, and $\boldsymbol{n}$ is the unit outward normal vector.

Figure~\ref{C_20} presents the results of drag and lift coefficients at $\mathrm{Re}=20$. Owing to the low Reynolds number, the flow is symmetric along the $x$ axis, and both methods exhibit stable lift coefficients near zero. Regarding the drag coefficient, as previously mentioned, the DDF method, due to the unsatisfaction of \eqref{conservation} in the force spreading process, results in a smaller drag coefficient. Conversely, the proposed method closely aligns with the results of Tafuni et al \cite{tafuni2018versatile}. Figure~\ref{C_200} shows the drag and lift coefficients at $\mathrm{Re}=200$. Similar to the results at $\mathrm{Re}=20$, the proposed method produces a drag coefficient that is close to the results of Thanh et al. \cite{bui2021simplified}, while the DDF method yields a lower drag coefficient. As for the lift coefficient, both results experience a minor disparity, with both being slightly lower than the results reported by Thanh et al. And their periodic behaviors remain similar.

\section{Conclusion}\label{Sec: conclusion}

This paper presents a novel moving lesat square immersed boundary method for smoothed particle hydrodynamics (SPH) with thin-walled structures.

By adopting the moving least square method for velocity interpolation, the method is stable for three-dimensional problems. Compared with traditional fixed/ghost particle wall boundary conditions, the immersed boundary method approach requires only a single layer of boundary particles, making it particularly well-suited for thin-walled structures immersed in fluid. The proposed approach is based on the direct forcing scheme, utilizing the moving lesat square method interpolation to obtain a smooth velocity field, ensuring stability in the results. Unlike methods based on the diffusive direct forcing approach, the proposed method avoids mutual interference on both sides of the thin-walled structure, resulting in a more accurate calculation of the force acting on the structure surface. Using interpolation instead of external extrapolation used in the moving least square diffusive direct forcing method, our approach avoids instability caused by insufficient support domain particles, making it stable for three-dimensional problems.


\section*{Acknowledgment}
This work has been supported by the National Natural Science Foundation of China (NSFC) under Grant No. 11972384, the Guangdong Basic and Applied Basic Research Foundation - Guangdong-Hong Kong-Macao Applied Mathematics Center Project under Grant No. 2021B1515310001, and the Guangdong Basic and Applied Basic Research Foundation - Regional Joint Fund Key Project under Grant No. 2022B1515120009. Additionally, we extend our appreciation to the National Key Research and Development Program under Grant No. 2020YFA0712502 for their invaluable support in this research.

\appendix
\section{Details of the Moving Least Square Interpolation} \label{appendix}

Consider a set of data points in a scalar field $\phi$. The linear interpolation value can be given by: 
\begin{equation}\label{polynomial}
		\widetilde{\phi}_j = c_1 x_{ij} + c_2 y_{ij} + c_3 z_{ij} + c_4,
\end{equation}
where subscript $j$ represents point index, $c_1,c_2,c_3$ are undetermined coefficient, $x_{ij}= x_i-x_j, y_{ij}=y_i-y_j, z_{ij}=z_i-z_j$ are the distance between point $j$ and center point $i$.

The weighted sum of squared differences between interpolation value and real value is: 
\begin{equation*}
	J = \sum_{j}^{N} W_{ij}^0(\widetilde{\phi}_j  - \phi_j)^2,
\end{equation*}
where $N$ is the number of data points. $W_{ij}^0$ is the weight, in Nasar's work$W_{ij}^0=W_{ij}V_j$, in our work $W_{ij}^0=W_{ij}$. Actually, as the volume fluctuations remain lower than $1\%$, the two weights have similar effects.
To minimize the error, we differentiate $J$ with respect to each coefficient $c_j$ and set the derivatives to zero:
\begin{equation*}
		\frac{\partial J}{\partial \boldsymbol{c}} = 0,
\end{equation*}
where $\boldsymbol{c}=\left[\begin{matrix}
	c_1 & c_2& c_3& c_4
\end{matrix}\right]^T$.

Solving these equations gives us the system of equations in matrix form as:
\begin{equation*}
	\sum_{i}^{N} W_{ij}^0
	\left[
		\begin{matrix}
			1	 &x_{ij}     &y_{ij}     &z_{ij}\\
			x_{ij}  &x_{ij}^2 &x_{ij} y_{ij} &x_{ij} z_{ij}\\
			y_{ij}  &y_{ij} x_{ij} &y_{ij}^2 &y_{ij} z_{ij} \\
			z_{ij}   &z_{ij} x_{ij} &z_{ij} y_{ij} &z_{ij}^2 \\
		\end{matrix}	
	\right]
	\left[
		\begin{matrix}
			c_1\\
			c_2\\
			c_3\\
			c_4\\
		\end{matrix}	
	\right]
	=\sum_{j}^{N} W_{ij}^0\phi_j
	\left[
	\begin{matrix}
		1\\
		x_{ij}\\
		y_{ij}\\
		z_{ij}\\
	\end{matrix}	
	\right]	.
\end{equation*}
\noindent
Solving this system of equations gives us the coefficients $c_1, c_2, c_3, c_4$ of the polynomial that best fits the data. Defining the matrix $A$:
\begin{equation*}
	A_i = 	\sum_j^N\left[
	\begin{matrix}
		1	 &x_{ij}     &y_{ij}     &z_{ij}\\
		x_{ij}  &x_{ij}^2 &x_{ij} y_{ij} &x_{ij} z_{ij}\\
		y_{ij}  &y_{ij} x_{ij} &y_{ij}^2 &y_{ij} z_{ij} \\
		z_{ij}   &z_{ij} x_{ij} &z_{ij} y_{ij} &z_{ij}^2 \\
	\end{matrix}	
	\right] W_{ij}^0,
\end{equation*}
then $\boldsymbol{c}$ can be represented as:
\begin{equation}\label{coefficient}
	\left[
	\begin{matrix}
		c_1\\
		c_2\\
		c_3\\
		c_4\\
	\end{matrix}	
	\right]
	=	A_i^{-1}
	\sum_{j}^{N} 
	\left[
\begin{matrix}
	1\\
	x_{ij}\\
	y_{ij}\\
	z_{ij}\\
\end{matrix}	
\right]	W_{ij}^0\phi_j .
\end{equation}

Substitute \eqref{coefficient} into \eqref{polynomial}, considering that for the center point $i$, $x_{ii} = 0, y_{ii}=0, z_{ii}=0$, we gets:
\begin{equation*}
	\widetilde{\phi}_i = 
	\left[
	\begin{matrix}
		1 &0 &0 &0
	\end{matrix}
	\right]
	\left[
	\begin{matrix}
		c_1\\
		c_2\\
		c_3\\
		c_4\\
	\end{matrix}	
	\right]
	= \sum_{j}^{N}
	 \left(
	 	\left[
	 \begin{matrix}
	 	1 &0 &0 &0
	 \end{matrix}
	 \right]
A_b^{-1}
	\left[
\begin{matrix}
	1\\
	x_{ij}\\
	y_{ij}\\
	z_{ij}\\
\end{matrix}	
\right]	W_{ij}^0\phi_j  
	\right)=\sum_{i}^{N} W_{ij}^{\mathrm{MLS}}\phi_j  ,
\end{equation*}
where
\begin{equation*}
	W_{ij}^{\mathrm{MLS}} = 
\left[
\begin{matrix}
	1 &0 &0 &0
\end{matrix}
\right]
A_b^{-1}
\left[
\begin{matrix}
	1\\
	x_{ij}\\
	y_{ij}\\
	z_{ij}\\
\end{matrix}	
\right]	W_{ij}^0,
\end{equation*}
the same as \eqref{W3d}. It can easily understood that in 2-D space, it becomes \eqref{Wmls}.

\bibliographystyle{elsarticle-num}
\bibliography{ref}
\end{document}